\documentclass[11pt,a4paper]{article}

\usepackage[T1]{fontenc}
\usepackage[utf8]{inputenc}
\usepackage{lmodern}
\usepackage{microtype}
\usepackage[margin=2.25cm]{geometry}
\usepackage{amsmath,amssymb,mathtools,bm}
\usepackage{booktabs,tabularx,array,multirow}
\usepackage{enumitem}
\usepackage{setspace}
\usepackage{xcolor}
\usepackage{hyperref}
\usepackage[nameinlink,capitalise,noabbrev]{cleveref}
\usepackage{titlesec}
\usepackage[backend=biber, style=numeric-comp, sorting=none,
    sortcites=true, maxbibnames=99, backref=true]{biblatex} 
\usepackage{comment}

\definecolor{linkblue}{RGB}{20,65,120}
\hypersetup{
  colorlinks=true,
  linkcolor=linkblue,
  citecolor=linkblue,
  urlcolor=linkblue,
  pdftitle={From Nonmetricity Operators to the Physical Spectrum},
  pdfauthor={QgravSTP manuscript draft}
}

\setlist[itemize]{leftmargin=1.6em,itemsep=1pt,topsep=3pt}
\setlist[enumerate]{leftmargin=1.8em,itemsep=1pt,topsep=3pt}
\allowdisplaybreaks
\titleformat{\section}{\large\bfseries}{\thesection.}{0.55em}{}
\titleformat{\subsection}{\normalsize\bfseries}{\thesubsection.}{0.5em}{}
\titleformat{\subsubsection}{\normalsize\itshape}{\thesubsubsection.}{0.5em}{}

\newcommand{\Q}{\mathbb Q}
\newcommand{\K}{\mathcal K}
\newcommand{\D}{\mathcal D}
\newcommand{\A}{\mathcal A}
\newcommand{\Lag}{\mathcal L}
\newcommand{\dd}{\mathrm d}
\newcommand{\Ptwo}{\mathbb{P}^{(2)}}
\newcommand{\Pone}{\mathbb{P}^{(1)}}
\newcommand{\Pzs}{\mathbb{P}^{(0\mathrm{s})}}
\newcommand{\Pzw}{\mathbb{P}^{(0\mathrm{w})}}
\newcommand{\Pzsw}{\mathbb{P}^{(0\mathrm{sw})}}
\newcommand{\Pzws}{\mathbb{P}^{(0\mathrm{ws})}}

\newcommand{\DOF}{\mathrm{DOF}}
\newcommand{\Diff}{\mathrm{Diff}}
\newcommand{\TDiff}{\mathrm{TDiff}}
\newcommand{\WTDiff}{\mathrm{WTDiff}}

\begin{document}

\begin{center}
{\LARGE\bfseries From Nonmetricity Operators to the Physical Spectrum:\\[3pt]
A Branch-Complete Analysis of Local Four-Derivative\\[3pt]
Symmetric Teleparallel Gravity}\par
\vspace{0.5cm}
{\large Caglar Pala}\par
\vspace{0.1cm}
{\small Department of Physics, Erciyes University}\par
\vspace{0.1cm}
{\small caglar.pala@gmail.com}\par
\end{center}

\begin{abstract}
We develop a branch-complete classical free-spectrum analysis of a local parity-even symmetric teleparallel theory containing all pure-gravity nonmetricity operators through four derivatives.  After algebraic canonicalisation and integration-by-parts reduction, the nonlinear action contains 116 independent operators distributed as $5+13+29+69$ among the sectors $Q^2$, $(\nabla Q)^2$, $Q^2\nabla Q$ and $Q^4$.  Around Minkowski spacetime in the coincident gauge only the first two sectors enter the quadratic action, which reduces exactly to four two-derivative and five four-derivative bilinears.  We give the explicit map from the original operator coefficients to this nine-dimensional basis, construct the Hessian and momentum kernel, classify the local linear gauge symmetries and show that unrestricted linearised diffeomorphism invariance imposes six independent relations leaving a three-parameter family $(A,B,C)$. Beyond the unrestricted Diff locus, we perform a systematic classification of reduced and accidental local linear symmetries within the certified vector--scalar generator module. The analysis recovers the TDiff, Weyl and WTDiff branches, establishes \(\mathcal S_{\mathrm{WTDiff}}=\mathcal S_{\mathrm{TDiff}}\cap\mathcal S_{\mathrm{Weyl}}\), and identifies additional accidental symmetry branches. Barnes--Rivers decomposition, covariant gauge fixing, exact inversion and conserved-source saturation yield
\begin{align*}
\A(q)=
\frac{3I_1-T^2}{3q(A+2Cq)}
-\frac{T^2}{3q\,[2A+(3B-2C)q]}
\end{align*}
The generic regular spectrum contains a massless graviton, a massive spin-two state and a massive scalar, with eight degrees of freedom in total.  The massive spin-two residue is necessarily opposite to the healthy massless graviton residue.  The complete regular branch classification leaves only the GR/STEGR locus $A>0$ with $B=C=0$ and the scalar extension $A>0$ with $B>0$ and $C=0$ as fully healthy Minkowski loci for a conserved metric source.  The latter has three degrees of freedom and mixed ultraviolet behaviour, with tensor exchange scaling as $k^{-2}$ and scalar exchange as $k^{-4}$. We place these results within the established symmetric-teleparallel, curvature-squared and enlarged metric-affine literature. Within this finite local metric theory, \(k^{-4}\) suppression of the source-coupled tensor channel requires the finite massive spin-two pole with opposite residue, so a healthy regular spin-two spectrum and full four-derivative tensor suppression are mutually exclusive.
\end{abstract}


\section{Introduction}

General relativity (GR) remains the uniquely successful classical description of gravitation over an enormous range of scales.  Its geometric formulation identifies gravity with the curvature of a Lorentzian metric and packages the dynamics into the Einstein--Hilbert action, whose equations are second order and whose linearisation around Minkowski spacetime propagates only the two helicities of a massless spin-two particle \cite{Einstein1916}.  At the same time, the quantum status of the theory has been problematic from the beginning of perturbative quantum gravity.  The dimensionful Newton coupling makes the Einstein theory nonrenormalisable by conventional power counting, while the counterterm structure generated by loops requires higher-curvature operators that are absent from the classical action.  The local four-derivative completion studied by Stelle is power-counting renormalisable, but its improved propagator comes with a massive spin-two pole of opposite residue \cite{Stelle1977,Stelle1978}.  This tension between ultraviolet improvement and perturbative unitarity has shaped essentially every later attempt to construct a local quantum theory of the metric.

The classical spectrum is therefore not merely a preliminary calculation.  It is the first consistency gate for any candidate quantum-gravity action.  Before discussing loops, renormalisation-group flows or ultraviolet fixed points, one must know which fields propagate, which poles are physical, which residues have the correct orientation, where tachyons occur and which parameter limits alter the rank of the kinetic operator.  The familiar curvature-squared examples illustrate why this must be done branch by branch.  The $R+R^2$ theory removes the massive spin-two pole and retains a healthy scalar, a mechanism that underlies Starobinsky inflation \cite{Starobinsky1980}.  Einstein--Weyl gravity removes the scalar but retains the ghostlike massive spin-two state.  Pure quadratic theories, critical limits and pole collisions are still more delicate because a limiting propagator need not describe the same constrained system as the action obtained after setting a kinetic coefficient to zero.  Modern studies and reviews continue to emphasise that power-counting renormalisability, absence of tachyons, absence of ghosts and nonlinear stability are logically distinct requirements \cite{Salvio2018,LuPerkinsPopeStelle2015,LuPope2011}.

A second motivation comes from the geometry used to represent gravity.  The Levi--Civita connection is fixed by metric compatibility and vanishing torsion, but it is not the only natural gravitational connection.  Gauge approaches initiated by Utiyama and Kibble promoted spacetime symmetries to local symmetries and led to theories with independent connection variables, torsion and eventually nonmetricity \cite{Utiyama1956,Kibble1961}.  Metric-affine gravity (MAG) treats the metric and affine connection as independent fields and therefore accommodates curvature, torsion and nonmetricity simultaneously.  Its kinematics is broad enough to couple not only to energy--momentum but also to spin, dilation and shear currents; the corresponding framework and Noether identities were systematised in the classic review of Hehl, McCrea, Mielke and Ne'eman \cite{Hehl1995}.  This enlarged field content can change the physical meaning of higher derivatives.  A curvature-squared term that is fourth order in a metric-only theory can be only second order in an independent connection.  Consequently, the occurrence of a healthy massive spin-two particle in an enlarged geometric model does not by itself contradict the standard metric higher-derivative ghost result.

Symmetric teleparallel geometry occupies a particularly instructive corner of this landscape.  It imposes
\begin{align}
R^\alpha{}_{\beta\mu\nu}(\Gamma)=0
\qquad\qquad
T^\alpha{}_{\mu\nu}(\Gamma)=0
\qquad\qquad
Q_{\alpha\mu\nu}:=\nabla_\alpha g_{\mu\nu}\neq0
\end{align}
so that gravitation is encoded in nonmetricity rather than curvature or torsion.  The flat and torsionless connection is locally pure gauge and can be set to zero in the coincident gauge.  The symmetric teleparallel equivalent of GR (STEGR) was developed as a dynamically equivalent reformulation of Einstein gravity \cite{NesterYo1999}, and the modern ``coincident GR'' formulation made its simplicity and covariance especially transparent \cite{CoincidentGR}.  Together with the curvature and torsion formulations, it forms the geometrical trinity of equivalent descriptions of GR \cite{GeometricalTrinity}.

There is also a substantial line of symmetric-teleparallel work by Adak, Dereli, Pala and collaborators that is directly relevant to the geometric setting adopted here.  Early studies reformulated GR and constructed exact solutions in curvature-free and torsion-free nonmetric geometry \cite{AdakSert2005,Adak2006}; a generic quadratic nonmetricity model was analysed in two dimensions by Adak and Dereli \cite{AdakDereli2008}; and a gauge-theoretic formulation of symmetric teleparallel gravity was developed by Adak \cite{Adak2018}.  More recent work has revisited coincident-gauge trajectories and autoparallels \cite{PalaAdak2022}, Weyl covariance and the second-clock question \cite{PalaSertAdak2023}, general teleparallel metric geometries and their gauge choices \cite{AdakDereliKoivistoPala2023}, and Weyl--Lorentz--$U(1)$ invariant symmetric teleparallel models \cite{AdakOzdemirPala2023}.  These works emphasise complementary geometric, variational and phenomenological aspects of nonmetricity gravity and provide useful context for the operator-level free-spectrum problem considered below.

Beyond the equivalent GR point, a generic nonmetricity action is a genuinely different theory: its gauge structure, constraints, perturbative spectrum and nonlinear degrees of freedom depend sensitively on the chosen operators and on whether the connection is retained as an independent constrained field.  The free spectrum of general analytic symmetric teleparallel metric theories was derived by Conroy and Koivisto using spin projectors and nonlocal form factors \cite{ConroyKoivisto2018}.  Related analyses classified teleparallel spectra and gravitational-wave polarisations \cite{KoivistoTsimperis2019,Hohmann2019}, while more recent work has sharpened the distinction between coincident-gauge descriptions, linearised metric symmetries and the full constraint problem of Newer GR and $f(Q)$ models \cite{Blixt2024,Zhao2024,Ferrara2026,JimenezHeisenbergKoivistoPekar2020,Hu2022,Dambrosio2023,Gomes2024,HuYamakoshi2023}.  The lesson is that no universal degree-of-freedom count can be assigned to all quadratic nonmetricity theories without specifying field content, background, symmetry branch and constraint rank.

The present work addresses a narrower but structurally complete problem.  We begin from the full local parity-even pure-gravity symmetric teleparallel operator basis through four derivatives rather than postulating a propagator-level ansatz.  The basis contains all contractions in the sectors $Q^2$, $(\nabla Q)^2$, $Q^2\nabla Q$ and $Q^4$, reduced under tensor symmetries and integration by parts.  We then follow the complete chain
\begin{align*}
\boxed{\begin{gathered}
116\ \text{nonlinear operators}
\longrightarrow 9\ \text{quadratic bilinears}
\longrightarrow \text{momentum kernel }\K(k)\\
\longrightarrow \text{Diff locus with 3 independent parameters}
\longrightarrow \text{propagator/exchange amplitude }\A(q)\\
\longrightarrow \text{branches, modes and health}
\end{gathered}}
\end{align*}
The emphasis is not on claiming the first higher-derivative symmetric teleparallel propagator.  Rather, it is on connecting a complete finite local nonlinear hierarchy to an exact free operator, a certified symmetry classification, a branch-complete spectrum and a parameter-space health theorem.  This supplies the operator-to-spectrum bridge needed before the cubic and quartic interactions are used in a quantum analysis.

The comparison with established theories is incorporated throughout rather than isolated in a separate section.  The physical kernel obtained below maps exactly to the affine-polynomial specialisation of the Conroy--Koivisto form factors \cite{ConroyKoivisto2018}.  The regular branches reproduce the familiar GR, $R+R^2$, Einstein--Weyl and generic quadratic-curvature spectra after convention translation \cite{Stelle1977,Starobinsky1980,Salvio2018}.  Analytic $f(Q)$ gravity around a regular $Q=0$ Minkowski vacuum is GR-like at quadratic order, whereas additional modes discussed in Hamiltonian and cosmological analyses concern nonlinear, background-dependent or strongly coupled sectors \cite{JimenezHeisenbergKoivistoPekar2020,Hu2022,Dambrosio2023,Gomes2024,HuYamakoshi2023}.  Finally, healthy extra particles in metric-affine and Riemann--Cartan models arise in enlarged field spaces with independent connection or torsion modes \cite{BaldazziMelichevPercacci2022,PercacciSezgin2020,MikuraPercacci2025,Percacci2026,AokiMukohyama2019,Neville1978,SezginVanNieuwenhuizen1980,Sezgin1981,LinHobsonLasenby2019,Marzo2022}; they are therefore comparison points rather than direct parameter points of the metric-only kernel studied here.

Our main conclusions can be stated compactly.  On the unrestricted linearised-diffeomorphism locus, the conserved-source exchange is governed by only two physical polynomial factors, $A+2Cq$ in the spin-two sector and $2A+(3B-2C)q$ in the scalar sector.  Generically they generate a massless graviton, a massive spin-two particle and a massive scalar.  The spin-two partial fraction forces the massive residue to oppose the massless graviton residue independently of the scalar sector.  The exact classification then leaves only two fully healthy regular Minkowski loci: GR/STEGR and a spin-two-reduced scalar branch.  Full $k^{-4}$ falloff in both source-coupled channels occurs only when the ghostlike massive spin-two pole is present, whereas the healthy scalar branch has mixed $k^{-2}/k^{-4}$ behaviour.  Singular $A=0$ and identically vanishing kinetic blocks are kept separate from ordinary particle branches rather than assigned unsupported degree-of-freedom counts.

The paper is organised as follows.  \Cref{sec:geometry} introduces the geometric setting and complete local operator hierarchy.  \Cref{sec:quadratic} derives the nine-dimensional quadratic action, the exact coefficient map, the Ward constraints and the reduced-symmetry atlas.  \Cref{sec:propagator} performs the Barnes--Rivers change of basis, gauge fixing, exact inversion and conserved-source saturation.  \Cref{sec:spectrum} derives the poles, residues and mode counts.  \Cref{sec:classification} gives the branch-complete parameter classification and the two healthy regular loci.  We conclude in \Cref{sec:conclusion} by collecting the physical results, their relation to the relevant literature and the questions deferred to nonlinear and quantum analysis.

\section{Geometric setting and complete local operator hierarchy}
\label{sec:geometry}

\subsection{Symmetric teleparallel variables}

We work in four dimensions with metric signature $(-,+,+,+)$.  The affine connection is constrained to be flat and torsionless, while nonmetricity is allowed.  Its two traces are
\begin{align}
Q_\alpha:=Q_{\alpha\mu}{}^{\mu}
\qquad
\widetilde Q_\alpha:=Q_{\mu\alpha}{}^{\mu}
\end{align}
A conventional STEGR scalar is
\begin{align}
\Q={}&-\frac14 Q_{\alpha\mu\nu}Q^{\alpha\mu\nu}
+\frac12 Q_{\alpha\mu\nu}Q^{\mu\alpha\nu}
+\frac14 Q_\alpha Q^\alpha
-\frac12 Q_\alpha\widetilde Q^\alpha
\label{eq:Qscalar}
\end{align}
Up to the sign convention for $Q_{\alpha\mu\nu}$, the Levi--Civita Ricci scalar satisfies
\begin{align}
\mathring R=-\Q+\mathring\nabla_\mu\bigl(Q^\mu-\widetilde Q^\mu\bigr)
\label{eq:boundary}
\end{align}
which explains the dynamical equivalence of the STEGR action and the Einstein--Hilbert action under standard boundary conditions \cite{NesterYo1999,CoincidentGR}.

The most general parity-even scalar quadratic in nonmetricity is spanned by five contractions,
\begin{align}
\Lag_{Q^2}={}&a_1 Q_{\alpha\mu\nu}Q^{\alpha\mu\nu}
+a_2 Q_{\alpha\mu\nu}Q^{\mu\alpha\nu}
+a_3 Q_\alpha Q^\alpha
+a_4 Q_\alpha\widetilde Q^\alpha
+a_5 \widetilde Q_\alpha\widetilde Q^\alpha
\label{eq:fiveQ}
\end{align}
The STEGR point is one special linear combination.  Away from it, the five parameters describe the quadratic-nonmetricity family usually associated with Newer GR, although direct comparison requires attention to coefficient ordering, overall signs, and whether the inertial connection perturbation is retained.

\subsection{Local four-derivative action}

The theory studied here contains all local parity-even pure-gravity contractions made from $Q_{\alpha\mu\nu}$ and its derivatives with at most four derivatives of the metric.  After canonicalisation under intrinsic index symmetries, exchange of identical factors, dummy-index relabelling, and integration by parts, the independent sectors are
\begin{align}
\mathcal S_1 &: Q^2 & \dim\mathcal S_1&=5\nonumber\\
\mathcal S_2 &: (\nabla Q)^2 & \dim\mathcal S_2&=13\nonumber\\
\mathcal S_3 &: Q^2\nabla Q & \dim\mathcal S_3&=29\nonumber\\
\mathcal S_4 &: Q^4 & \dim\mathcal S_4&=69
\end{align}
Thus the nonlinear action contains 116 independent operators,
\begin{align}
S=\int\dd^4x\sqrt{-g}\left(
\sum_{i=1}^{5}a_i\mathcal O_i^{(Q^2)}
+\sum_{i=1}^{13}b_i\mathcal O_i^{((\nabla Q)^2)}
+\sum_{i=1}^{29}c_i\mathcal O_i^{(Q^2\nabla Q)}
+\sum_{i=1}^{69}d_i\mathcal O_i^{(Q^4)}
\right)
\label{eq:fullaction}
\end{align}
This basis is defined in generic dimension before any four-dimensional curvature identities are invoked.  It is therefore adapted to a systematic counterterm analysis: the free theory is a projection of the same operator hierarchy that later supplies cubic and quartic vertices.

\subsection{Minkowski expansion and perturbative order}

We expand
\begin{align}
g_{\mu\nu}=\eta_{\mu\nu}+h_{\mu\nu}
\end{align}
and use the coincident representative $\Gamma^\alpha{}_{\mu\nu}=0$.  Then
\begin{align}
Q_{\alpha\mu\nu}=\partial_\alpha h_{\mu\nu}
\qquad\qquad
\nabla_\rho Q_{\alpha\mu\nu}=\partial_\rho\partial_\alpha h_{\mu\nu}
\end{align}
Consequently,
\begin{align}
Q^2&=O(h^2) & (\nabla Q)^2&=O(h^2)
&Q^2\nabla Q&=O(h^3) & Q^4&=O(h^4)
\label{eq:counting}
\end{align}
Only $\mathcal S_1$ and $\mathcal S_2$ contribute to the free action. Accordingly, Appendix \ref{app:explicit forms of operators} lists explicitly only the \(5+13\) operators belonging to the \(Q^2\) and \((\nabla Q)^2\) sectors that enter the present quadratic analysis; the remaining \(29+69\) interaction operators in \(Q^2\nabla Q\) and \(Q^4\) are not reproduced here since they first contribute at cubic and quartic order, respectively. This observation sharply separates two tasks that are often conflated.  The first two sectors determine the classical propagator, the latter two do not alter the quadratic spectrum but are essential for interactions, loop divergences, and radiative stability.  A healthy free-theory locus is therefore not automatically stable under quantum corrections unless the nonlinear couplings enforce the same restriction on counterterms.

\section{Quadratic action, kernel, and gauge structure}
\label{sec:quadratic}

The free quadratic action receives contributions only from $\mathcal S_1\propto Q^2$ and $\mathcal S_2\propto(\nabla Q)^2$
\begin{align}
S^{(2)} = \int d^4x\sqrt{-\eta}\,\Lag^{(2)}
\end{align}
where
\begin{align}
    \Lag^{(2)}=
\sum_{i=1}^{5}a_i\mathcal O_i^{(Q^2)}
+\sum_{i=1}^{13}b_i\mathcal O_i^{((\nabla Q)^2)}
\label{eq:Lagrangian original free theory 18 terms}
\end{align}
Here $\Lag^{(2)}$ and $S^{(2)}$ denote the Lagrangian and action at quadratic order in the metric perturbation $h$.

\subsection{Reduction to nine bilinears}

After integration by parts, the complete quadratic action reduces to nine independent bilinears,
\begin{align}
\Lag^{(2)}=
\sum_{i=1}^{4}\alpha_i\mathcal B_i^{(2)}
+\sum_{i=1}^{5}\beta_i\mathcal B_i^{(4)}
\label{eq:nineaction}
\end{align}
Here $\mathcal{B}_i^{(2)}$ and $\mathcal{B}_i^{(4)}$ denote bilinears with two and four derivatives, respectively. A convenient tensor-family representation is obtained from the trace $h=\eta^{\mu\nu}h_{\mu\nu}$, the divergence $D_\nu=\partial^\mu h_{\mu\nu}$, $X=\partial^\mu\partial^\nu h_{\mu\nu}$ and $\Box=\partial_\mu\partial^\mu$

\begin{subequations}
\begin{align}
    \mathcal{B}_1^{(2)} &= \partial_\mu h D^\mu &\qquad \mathcal{B}_1^{(4)} &= -\partial_\mu h\Box D^\mu \\
    \mathcal{B}_2^{(2)} &= \partial_\mu h \partial^\mu h &\qquad \mathcal{B}_2^{(4)} &= h\Box^2 h \\
    \mathcal{B}_3^{(2)} &= \partial_\rho h_{\mu\nu} \partial^\rho h^{\mu\nu} &\qquad \mathcal{B}_3^{(4)} &= h_{\mu\nu}\Box^2 h^{\mu\nu} \\
    \mathcal{B}_4^{(2)} &= D_\mu D^\mu &\qquad \mathcal{B}_4^{(4)} &= -D_\mu\Box D^\mu \\
    & &\qquad \mathcal{B}_5^{(4)} &= X^2
    \label{eq:schematicbilinears}
\end{align}    
\end{subequations}
where equality is understood at the action level modulo total derivatives.  The minus signs in $\mathcal B_1^{(4)}$ and $\mathcal B_4^{(4)}$ are fixed by the explicit $(\nabla Q)^2$ operators in Appendix \ref{app:explicit forms of operators} and ensure that the reduced basis, Hessian signs, and coefficient dictionary are mutually consistent.  The relation between the coefficients of the original quadratic action $\{a_i,b_i\}$ and the reduced bilinears $\{\alpha_i,\beta_i\}$ may be given as follows
\begin{subequations}
\begin{align}
    \alpha_1 &= a_4 &\qquad \beta_1 &= b_2+b_3+b_7 \\
    \alpha_2 &= a_3 &\qquad \beta_2 &= b_1+b_{10} \\
    \alpha_3 &= a_1 &\qquad \beta_3 &= b_5+b_{11} \\
    \alpha_4 &= a_2+a_5 &\qquad \beta_4 &= b_4+b_8+b_{12} \\
    & &\qquad \beta_5 &= b_6+b_9+b_{13}
\end{align}    
\end{subequations}
The exact construction was performed by varying every canonical bilinear, moving derivatives by deterministic integration by parts, and assembling a formally self-adjoint Hessian,
\begin{align}
S^{(2)}=\frac12\int\dd^4x\,
 h_{\mu\nu}\K^{\mu\nu,\rho\sigma}(\partial)h_{\rho\sigma}
\label{eq:hessianform}
\end{align}
where the explicit form of the Hessian expanded in the canonical (Euler) bases reads
\begin{align}
    \K^{\mu\nu,\rho\sigma}(\partial) &= -\alpha_1 \mathbb{H}_1^{(2)} -2\alpha_2 \mathbb{H}_2^{(2)} -2\alpha_3 \mathbb{H}_3^{(2)} - 2\alpha_4 \mathbb{H}_4^{(2)} \nonumber \\
    &\quad +\beta_1 \mathbb{H}_1^{(4)} +2\beta_2 \mathbb{H}_2^{(4)} + 2\beta_3 \mathbb{H}_3^{(4)} + 2\beta_4 \mathbb{H}_4^{(4)} + 2\beta_5 \mathbb{H}_5^{(4)}
\end{align}
Here the Euler bases are given as follows
\begin{subequations}
    \begin{align}
    \mathbb{H}_1^{(2)} &= \eta^{\rho\sigma}\,\partial^{\mu}\,\partial^{\nu} + \eta^{\mu\nu}\,\partial^{\rho}\,\partial^{\sigma} &\quad \mathbb{H}_1^{(4)} &= (\eta^{\rho\sigma}\,\partial^{\mu}\,\partial^{\nu} + \eta^{\mu\nu}\,\partial^{\rho}\,\partial^{\sigma})\Box \\
\mathbb{H}_2^{(2)} &= \eta^{\mu\nu}\,\eta^{\rho\sigma}\Box &\quad \mathbb{H}_2^{(4)} &= \eta^{\mu\nu}\,\eta^{\rho\sigma}\Box^2 \\
\mathbb{H}_3^{(2)} &= I^{\mu\nu,\rho\sigma}\Box &\quad \mathbb{H}_3^{(4)} &= I^{\mu\nu,\rho\sigma}\Box^2 \\
\mathbb{H}_4^{(2)} &= \eta^{(\mu(\rho}\,\partial^{\nu)}\,\partial^{\sigma)} &\quad \mathbb{H}_4^{(4)} &= \eta^{(\mu(\rho}\,\partial^{\nu)}\,\partial^{\sigma)}\Box \\
& &\quad \mathbb{H}_5^{(4)} &= \partial^{\mu}\,\partial^{\nu}\,\partial^{\rho}\,\partial^{\sigma}
\end{align}
\end{subequations}
where the symmetric identity tensor is defined as follows
\begin{align}
I^{\mu\nu,\rho\sigma}:=\eta^{\mu(\rho}\eta^{\sigma)\nu}=\frac12\left(\eta^{\mu\rho}\eta^{\nu\sigma}+\eta^{\mu\sigma}\eta^{\nu\rho}\right)
\end{align}
It is useful to emphasize the relation between the quadratic and quartic bases
\begin{align}
    \mathbb{H}_i^{(4)} = \Box\,\mathbb{H}_i^{(2)} \qquad\qquad i=\{1,2,3,4\}
\end{align}
The independent first variation gives the Euler tensor
\begin{align}
\mathcal E^{\mu\nu}=\K^{\mu\nu,\rho\sigma}(\partial)\,h_{\rho\sigma}
\end{align}
and the linearised field equations are $\mathcal{E}^{\mu\nu}=0$. In a pure metric theory linearised diffeomorphism symmetry implies the off-shell Noether identity
\begin{align}
    \partial_\mu \mathcal{E}^{\mu\nu} = 0 \label{eq:Nother id}
\end{align}

\subsection{Momentum kernel and the exact diffeomorphism (Diff) locus}

We use Fourier transformation
\begin{align}
h(x)=\int\frac{\dd^4k}{(2\pi)^4}e^{ik\cdot x}h(k)
\qquad\qquad
\partial_\mu\mapsto ik_\mu
\qquad\qquad
q:=k^2
\end{align}
The quadratic action becomes
\begin{align}
S^{(2)}=\frac12\int\frac{\dd^4k}{(2\pi)^4}
 h_{\mu\nu}(-k)\K^{\mu\nu,\rho\sigma}(k)h_{\rho\sigma}(k)
\end{align}
Here the \textit{momentum kernel} $\K^{\mu\nu,\rho\sigma}(k)$ may be expanded in the canonical basis as follows
\begin{align}
    \K^{\mu\nu,\rho\sigma}(k) = \sum_{i=1}^5 m_i(q)\,\mathbb{K}_i
    \label{eq:kernel generic Euler bases}
\end{align}
where the coefficients and the bases read
\begin{subequations}
    \begin{align}
    m_1(q) &= \alpha_1 + \beta_1q &\qquad \mathbb{K}_1 &= \eta^{\rho\sigma} k^\mu k^\nu + \eta^{\mu\nu} k^\rho k^\sigma \\
    m_2(q) &= 2(\alpha_2 + \beta_2q)q &\qquad \mathbb{K}_2 & = \eta^{\mu\nu}\eta^{\rho\sigma} \\
    m_3(q) &= 2(\alpha_3 + \beta_3q)q &\qquad \mathbb{K}_3 &= I^{\mu\nu,\rho\sigma} \\
    m_4(q) &= 2(\alpha_4 + \beta_4q) &\qquad \mathbb{K}_4 &= \eta^{(\mu(\rho} k^{\nu)}k^{\sigma)} \\
    m_5(q) &= 2\beta_5 &\qquad \mathbb{K}_5 &= k^\mu k^\nu k^\rho k^\sigma
\end{align}
\end{subequations}
The infinitesimal diffeomorphism $x\mapsto x+\xi(x) $ induces the linearised metric transformation
\begin{align}
\delta_\xi h_{\mu\nu}=2\partial_{(\mu}\xi_{\nu)}
\label{eq:linDiff}
\end{align}
which we use below as the candidate local gauge transformation of the quadratic metric theory. Correspondingly, linearised diffeomorphism invariance is equivalent to the Ward identity
\begin{align}
k_\mu\K^{\mu\nu,\rho\sigma}(k)=0
\label{eq:ward}
\end{align}
This identity is the momentum-space counterpart of the position-space Noether identity given by Eq.\eqref{eq:Nother id}. The contraction separates into three independent tensor structures at each derivative order.  Requiring their coefficients to vanish yields six conditions,
\begin{subequations}
    \begin{align}
    \alpha_1+2\alpha_2&=0
    &\alpha_3+\frac12\alpha_4&=0
    &\alpha_1+\alpha_4&=0 \\
    \beta_1+2\beta_2&=0
    &\beta_3+\frac12\beta_4&=0
    &\beta_1+\beta_4+2\beta_5&=0
    \label{eq:wardconstraints}
    \end{align}
\end{subequations}
The rank of the constraint system is six, so the unrestricted $\Diff$ locus is three-dimensional.  The same equations follow independently from direct variation of the Lagrangian given by Eq.\eqref{eq:nineaction} and from the off-shell Noether identity given by Eq.\eqref{eq:Nother id}.

We denote the three independent combinations on this locus by $\{A,B,C\}$
\begin{subequations}
\begin{align}
    \alpha_1 &= A &\qquad \beta_1 &= B \\
    \alpha_2 &= -\frac{A}{2} &\qquad \beta_2 &= -\frac{B}{2} \\
    \alpha_3 &= \frac{A}{2} &\qquad \beta_3 &= C \\
    \alpha_4 &= -A &\qquad \beta_4 &= -2C \\
    & &\qquad \beta_5 &= -\frac{B-2C}{2}
\end{align}    
\end{subequations}
Their definition is most transparent through the physical spin coefficients in \Cref{sec:propagator}.  They should be understood as reduced free-theory parameters through $\Diff$ symmetry, not as a universal raw coefficient dictionary for every action in the literature.

\subsection{Reduced and accidental symmetry atlas}\label{sec:reduced symmetries}

The Diff locus is only one member of a larger quadratic symmetry atlas.  To display the structure compactly it is useful to introduce the momentum-space generator bases
\begin{subequations}
\begin{align}
V_1[\xi]_{\mu\nu}&=k_{(\mu}\xi_{\nu)}
&V_2[\xi]_{\mu\nu}&=\eta_{\mu\nu}(k\cdot\xi)\\
S_0[\phi]_{\mu\nu}&=\eta_{\mu\nu}\phi
&S_1[\phi]_{\mu\nu}&=k_\mu k_\nu\phi
&S_2[\phi]_{\mu\nu}&=q\eta_{\mu\nu}\phi
\end{align}    
\end{subequations}
where irrelevant overall normalisations of the gauge parameters have been suppressed. Then the most general local linear generator, within the five-generator module considered here, may be given in terms of the bases
\begin{align}
    G_{\mu\nu}[\xi,\phi] = g_1 V_1[\xi]_{\mu\nu} + g_2 V_2[\xi]_{\mu\nu} + g_3 S_0[\phi]_{\mu\nu} + g_4 S_1[\phi]_{\mu\nu} + g_5 S_2[\phi]_{\mu\nu}
\end{align}
and the familiar reduced symmetries follow directly from the kernel action on this generator
\begin{align}
    \K^{\mu\nu,\rho\sigma}(k)\,G_{\rho\sigma}[\xi,\phi] = 0
    \label{eq:generalised Ward identity}
\end{align}
 This is the \textit{generalised Ward identity}. The vector and scalar parameters are independent; Eq.\eqref{eq:generalised Ward identity} is imposed sectorwise, with the relevant generator coefficients activated branch by branch. Transverse diffeomorphisms (TDiff) require
\begin{align}
2\alpha_3+\alpha_4=0
\qquad\qquad
2\beta_3+\beta_4=0
\end{align}
and therefore define a rank-two, seven-dimensional coupling locus.  Weyl invariance requires the four independent trace conditions
\begin{subequations}
\begin{align}
\alpha_1+8\alpha_2+2\alpha_3&=0
&2\alpha_1+\alpha_4&=0\\
\beta_1+8\beta_2+2\beta_3&=0
&2\beta_1+\beta_4+\beta_5&=0
\end{align}    
\end{subequations}
which define a rank-four, five-dimensional locus.  Their simultaneous imposition gives Weyl-transverse diffeomorphisms (WTDiff)
\begin{align}
\mathcal S_{\WTDiff}=\mathcal S_{\TDiff}\cap\mathcal S_{\mathrm{Weyl}}
\qquad\qquad
\operatorname{rank}=6
\qquad\qquad
\dim\mathcal S_{\WTDiff}=3
\end{align}
with the explicit parametrisation
\begin{subequations}
\begin{align}
\alpha_1&=-\frac12\alpha_4
&\alpha_2&=\frac{3}{16}\alpha_4
&\alpha_3&=-\frac12\alpha_4\\
\beta_1&=-\frac12(\beta_4+\beta_5)
&\beta_2&=\frac{3\beta_4+\beta_5}{16}
&\beta_3&=-\frac12\beta_4
\end{align}    
\end{subequations}
The unrestricted generator search produces thirteen branches. The principal additional branches are represented in Appendix \ref{app:symmetry conditions}, while Table \ref{tab:symmetry-atlas} records the complete branch structure. Rank and nullity refer to the homogeneous constraint system in the nine-dimensional coupling space and must not be confused with a propagating degree-of-freedom count.

Two intersections are especially useful for interpreting the degeneracies.  Diff plus an independent Weyl symmetry has rank eight and leaves the one-dimensional coupling locus
\begin{align}
\alpha_1=\alpha_2=\alpha_3=\alpha_4=0
\qquad\qquad
\beta_1=\beta_5
\qquad\qquad
\beta_2=-\frac12\beta_5
\qquad\qquad
\beta_3=\frac32\beta_5
\qquad\qquad
\beta_4=-3\beta_5
\end{align}
which annihilates spin one and the complete scalar block at generic nonzero momentum.  By contrast, Diff already contains the unrestricted double-gradient image because a longitudinal parameter $\xi_\mu=k_\mu\phi$ gives $V_1[\xi]=S_1[\phi]$.  These results identify precisely which projector sectors become gauge-degenerate and therefore where gauge fixing or quotient-space inversion must precede the propagator construction.  The physical spectrum analysed below is intentionally restricted to the unrestricted Diff locus.

\begin{table}[!h]
\centering
\caption{Quadratic local symmetry atlas in the five-generator module.  The last two rows are restricted to the massless shell and do not represent unrestricted off-shell gauge symmetries.}
\label{tab:symmetry-atlas}
\small
\begin{tabularx}{\textwidth}{@{}>{\raggedright\arraybackslash}p{2.7cm}>{\raggedright\arraybackslash}p{3.1cm}>{\raggedright\arraybackslash}p{2.25cm}c>{\raggedright\arraybackslash}X@{}}
\toprule
Branch & Generator & Restriction & Rank/nullity & Interpretation \\
\midrule
Diff & $V_1$ & arbitrary vector & $6/3$ & full linearised diffeomorphism symmetry \\
Trace image & $V_2$ & arbitrary vector & $4/5$ & one-scalar image with the Weyl coupling ideal \\
Weyl & $S_0$ & arbitrary scalar & $4/5$ & unrestricted trace gauge symmetry \\
Double gradient & $S_1$ & arbitrary scalar & $4/5$ & genuine accidental scalar gauge branch \\
Metric-box image & $S_2=qS_0$ & arbitrary scalar & $4/5$ & polynomial image of the Weyl branch \\
Mixed vector & $V_1+rV_2$ & $r\neq-\tfrac14$ & $6/3$ & trace-deformed vector family with Diff recovered at $r=0$ \\
Mixed vector exceptional & $V_1-\tfrac14V_2$ & exceptional chart & $6/3$ & traceless longitudinal scalar component \\
Mixed scalar & $S_0+pS_1+tS_2$ & $p\neq0$ & $6/3$ & enhanced scalar locus annihilating the complete scalar block \\
Mixed scalar image & $S_0+tS_2$ & $p=0$ & $4/5$ & Weyl-equivalent polynomial multiple \\
Derivative scalar & $S_1+tS_2$ & arbitrary $t$ & $4/5$ & accidental one-scalar derivative family; $t=-\tfrac14$ requires a separate coordinate chart \\
TDiff & $V_1[\xi^{\mathrm T}]$ & $k\cdot\xi^{\mathrm T}=0$ & $2/7$ & transverse diffeomorphism symmetry \\
Harmonic Weyl & $S_0[\phi]$ & $q\phi=0$ & $1/8$ & restricted trace zero mode on the massless shell \\
Harmonic double gradient & $S_1[\phi]$ & $q\phi=0$ & $0/9$ & universal restricted null image on the massless shell \\
\bottomrule
\end{tabularx}
\end{table}

\section{Barnes--Rivers decomposition and the physical propagator}
\label{sec:propagator}

\subsection{Projector basis}

For non-null momentum define
\begin{align}
\theta_{\mu\nu}=\eta_{\mu\nu}-\frac{k_\mu k_\nu}{q}
\qquad\qquad
\omega_{\mu\nu}=\frac{k_\mu k_\nu}{q}
\end{align}
The symmetric rank-two identity decomposes into spin-two, spin-one, and scalar projectors \cite{Barnes1965,Rivers1964},
\begin{align}
I_{\mu\nu,\rho\sigma}
=\Ptwo+\Pone+\Pzs+\Pzw
\end{align}
with
\begin{align}
\Ptwo_{\mu\nu,\rho\sigma}
&=\frac12(\theta_{\mu\rho}\theta_{\nu\sigma}
+\theta_{\mu\sigma}\theta_{\nu\rho})
-\frac13\theta_{\mu\nu}\theta_{\rho\sigma}
\nonumber\\
\Pone_{\mu\nu,\rho\sigma}
&=\frac12(\theta_{\mu\rho}\omega_{\nu\sigma}
+\theta_{\mu\sigma}\omega_{\nu\rho}
+\theta_{\nu\rho}\omega_{\mu\sigma}
+\theta_{\nu\sigma}\omega_{\mu\rho})
\nonumber\\
\Pzs_{\mu\nu,\rho\sigma}
&=\frac13\theta_{\mu\nu}\theta_{\rho\sigma}
\qquad\qquad
\Pzw_{\mu\nu,\rho\sigma}=\omega_{\mu\nu}\omega_{\rho\sigma}
\end{align}
plus the mixed scalar matrix units
\begin{align}
\Pzsw=\frac1{\sqrt3}\theta_{\mu\nu}\omega_{\rho\sigma}
\qquad\qquad
\Pzws=\frac1{\sqrt3}\omega_{\mu\nu}\theta_{\rho\sigma}
\end{align}
The complete multiplication table reduces the inversion of a general quadratic kernel to one-dimensional spin-two and spin-one inversions plus a $2\times2$ scalar inverse.

Generic kernel given by Eq.\eqref{eq:kernel generic Euler bases} may be decomposed in the spin projector basis as follows
\begin{align}
    \K^{\mu\nu,\rho\sigma}(k) &= \sum_{i=1}^6 c_i(q)\,\mathbb{P}_i^{\mu\nu,\rho\sigma} \nonumber \\
    &= c_2(q)\Ptwo + c_1(q)\Pone + c_s(q)\Pzs + c_w(q)\Pzw + c_{sw}(q)\Pzsw + c_{ws}(q)\Pzws \label{eq:kernel generic BR bases}
\end{align}
where the coefficients read
\begin{subequations}
\begin{align}
    c_2(q) &= m_3(q) \\
    c_1(q) &= m_3(q) + q\frac{m_4(q)}{2} \\
    c_s(q) &= m_3(q) + 3m_2(q) \\
    c_w(q) &= 2qm_1(q)+m_2(q)+m_3(q)+qm_4(q)+q^2m_5(q) \\
    c_{sw}(q) &= \sqrt{3}[qm_1(q)+m_2(q)] \\
    c_{ws}(q) &= \sqrt{3}[qm_1(q)+m_2(q)]
\end{align}    
\end{subequations}
The Barnes--Rivers decomposition may therefore be viewed as a change of basis between the generic/Euler and spin-projector representations
\begin{align}
    m_i(q)\mathbb{K}_i \quad\mapsto\quad c_i(q)\mathbb{P}_i
\end{align}
which are given by Eq.\eqref{eq:kernel generic Euler bases} and Eq.\eqref{eq:kernel generic BR bases}, respectively.

\subsection{Physical kernel on the $\Diff$ locus}

On the Diff/Ward locus coefficients read
\begin{align}
    c_2 = q(A+2Cq) \qquad\qquad c_s = -q[2A+(3B-2C)q] \qquad\qquad c_1=c_w=c_{sw}=c_{ws}=0
\end{align}
and the ungauge-fixed physical kernel collapses to

\begin{align}
\K_{\Diff}(q)
=q(A+2Cq)\Ptwo
-q\bigl[2A+(3B-2C)q\bigr]\Pzs
\label{eq:physicalkernel}
\end{align}
The spin-one and longitudinal scalar directions are gauge null vectors.  A two-parameter covariant gauge-fixing term makes the full tensor operator invertible
\begin{align}
    \Lag_\mathrm{gf}=-\frac{1}{2Z_\mathrm{gf}}F_\mu F^\mu
\end{align}
where the generic covariant gauge family is defined by
\begin{align}
    F_\mu=\partial^\nu h_{\mu\nu} - Y_\mathrm{gf}\partial_\mu h
\end{align}
Here $\{Z_\mathrm{gf},Y_\mathrm{gf}\}$ are free parameters. One may choose the harmonic (de Donder) gauge by $Y_\mathrm{gf}=1/2$. Adding the gauge-fixing term gives the total kernel
\begin{align}
    \K_\mathrm{total}=\K_\Diff + \K_\mathrm{gf}
\end{align}
from which we derive the total propagator
\begin{align}
    \mathcal{D}_\mathrm{total} &= \K^{-1}_\mathrm{total} \nonumber \\
    &= \frac{1}{q(A+2Cq)}\Ptwo + \frac{2Z_\mathrm{gf}}{q}\Pone - \frac{1}{q[2A+(3B-2C)q]}\Pzs \nonumber \\
    &\quad + \frac{2AZ_\mathrm{gf} -3Y_\mathrm{gf}^2 + (3B-2C)Z_\mathrm{gf}q }{q(Y_\mathrm{gf}-1)^2[2A + (3B-2C)q]}\Pzw \nonumber \\
    &\quad + \frac{\sqrt{3}Y_\mathrm{gf}}{q(Y_\mathrm{gf}-1)[2A + (3B-2C)q]}(\Pzsw + \Pzws)
\end{align}
The Feynman propagator is $i\D$ in the adopted convention. Now, let $T_{\mu\nu}$ be a symmetric conserved source,
\begin{align}
k_\mu T^{\mu\nu}=0
\qquad\qquad
I_1:=T_{\mu\nu}T^{\mu\nu}
\qquad\qquad
T:=\eta^{\mu\nu}T_{\mu\nu}
\end{align}
The nontrivial elementary contractions are
\begin{align}
T\Ptwo T=I_1-\frac13T^2
\qquad\qquad
T\Pzs T=\frac13T^2
\end{align}
while all longitudinal and mixed contractions vanish
\begin{align}
    T\Pone T = T\Pzw T = T\Pzsw T = T\Pzws T = 0
\end{align}
In the total propagator, after saturation with a conserved source, every gauge-dependent longitudinal contribution vanishes. Hence the effective physical inverse is
\begin{align}
\D_{\mathrm{phys}}(q) =\frac{\Ptwo}{q(A+2Cq)}
-\frac{\Pzs}{q\,[2A+(3B-2C)q]}
\label{eq:physicalprop}
\end{align}
Correspondingly we may define the gauge-independent amplitude as follows
\begin{align}
\A(q):=T\D T
=\frac{3I_1-T^2}{3q(A+2Cq)}
-\frac{T^2}{3q\,[2A+(3B-2C)q]}
\label{eq:masteramplitude}
\end{align}
This is the master formula for the spectrum and parameter classification. Consequently, the source-visible pole locations and residues are strictly independent of the gauge-fixing parameters \(Z_{\rm gf}\) and \(Y_{\rm gf}\); these parameters occur only in longitudinal sectors that decouple from a conserved source.

\subsection{GR exchange as a consistency check}

Setting $B=C=0$ in Eq.\eqref{eq:masteramplitude} gives
\begin{align}
\A_{\mathrm{GR}}(q)
&=\frac{I_1-T^2/3}{Aq}-\frac{T^2}{6Aq}
=\frac1{Aq}\left(I_1-\frac12T^2\right)
\label{eq:GRamp}
\end{align}
The explicit scalar projector term in Eq.\eqref{eq:physicalprop} does not represent an independent scalar particle in GR.  It is part of the constrained massless spin-two exchange and combines with the transverse-traceless projector to produce the familiar conserved-source numerator.  An independent scalar degree of freedom requires a separate pole or a rank change, not merely the appearance of $\Pzs$ in a covariant propagator.

\section{Spectrum, residues, and degrees of freedom}
\label{sec:spectrum}

\subsection{Pole catalogue}
Define
\begin{align}
D_s:=3B-2C
\end{align}
For $A\neq0$, $C\neq0$, and $D_s\neq0$, the canonical pole locations are
\begin{align}
q_0=0
\qquad\qquad
q_2=-\frac{A}{2C}
\qquad\qquad
q_s=-\frac{2A}{D_s}
\label{eq:poles}
\end{align}
With the mostly-plus convention, a massive shell satisfies $q=-m^2$, so
\begin{align}
m_2^2=\frac{A}{2C}
\qquad\qquad
m_s^2=\frac{2A}{3B-2C}
\label{eq:masses}
\end{align}
The generic non-tachyon conditions are $A/C>0$ and $A/(3B-2C)>0$.

The pole factors already display the principal structural branches for $A\neq 0$. i) The limit $C=0$ removes the massive spin-two factor; ii) $D_s=0$ removes the massive scalar factor; iii) $B=C=0$ gives GR/STEGR; iv) the equality $B=2C$ makes the two masses equal when $A,C\neq0$.  The surfaces $A=0$ and the loci on which an entire physical kinetic block vanishes require separate treatment because the analytic order or rank of the operator changes.

\subsection{Residue theorem}
The spin-two factor admits the exact partial fraction
\begin{align}
\frac1{q(A+2Cq)}
=\frac1A\left(\frac1q-\frac1{q+m_2^2}\right)
\label{eq:spin2partial}
\end{align}
For $A>0$, the massless spin-two exchange has the healthy GR orientation. The massive spin-two pole necessarily has the opposite sign. This conclusion does not depend on $B$, on the scalar mass, or on whether the massive pole is tachyonic. It follows from the polynomial structure of a local fourth-order metric kinetic operator. Thus every regular branch with finite $C\neq0$ contains a massive spin-two ghost relative to a healthy massless graviton.

The scalar contribution similarly decomposes as
\begin{align}
-\frac1{q(2A+D_sq)}
=-\frac1{2A}\left(\frac1q-\frac1{q+m_s^2}\right)
\label{eq:scalarpartial}
\end{align}
Combining the massless pieces from Eq.\eqref{eq:spin2partial} and Eq.\eqref{eq:scalarpartial} reproduces Eq.\eqref{eq:GRamp}.  The massive scalar residue has the healthy orientation for $A>0$ once its kinetic normalisation and source structure are taken into account, while absence of a tachyon requires $m_s^2>0$.  In the spin-two-reduced branch this gives the simple conditions $A>0$ and $B>0$.

\subsection{Mode counting}
Independent projector-rank and explicit component analyses give
\begin{align}
\text{massless graviton:}&\quad 2\ \text{helicities}\nonumber\\
\text{massive spin-two:}&\quad 5\ \text{polarisations}\nonumber\\
\text{massive scalar:}&\quad 1\ \text{polarisation}
\end{align}
The generic regular theory therefore has
\begin{align}
N_{\DOF}=2+5+1=8
\end{align}
When $C=0$ the massive spin-two mode is removed and $N_{\DOF}=3$.  When $D_s=0$ the scalar is removed and $N_{\DOF}=7$.  GR/STEGR has two degrees of freedom.

The equal-mass condition is
\begin{align}
m_2^2=m_s^2
\quad\Longleftrightarrow\quad
B=2C
\qquad(A,C\neq0)
\label{eq:equalmass}
\end{align}
This does not produce a double pole.  The spin-two and scalar denominators become proportional, but they multiply orthogonal eigenspaces. The common shell therefore contains five spin-two and one scalar polarisation, each associated with a simple pole. This is an accidental equality of masses between two independent species, not an algebraic double pole. A repeated root of a single kinetic polynomial and two distinct spin projectors at equal mass are physically different notions.

\subsection{Critical and singular branches}
If $A=0$, the massless and massive roots collide at $q=0$ and the regular partial fractions above are invalid.  Depending on $B$ and $C$, the exchange may contain $q^{-2}$ dipole behaviour, proportional repeated factors, or identically vanishing kinetic blocks.  Likewise, if $A=C=0$ the spin-two operator is absent, while $A=D_s=0$ removes the scalar block. On the fully zero locus no physical quadratic operator exists. These cases cannot be assigned ordinary particle counts from a regular propagator limit alone. They may involve enhanced gauge symmetry, constraints, logarithmic or generalised modes, or strong coupling. We therefore quarantine them from the regular health theorem. This conservative treatment is also necessary when comparing with pure $R^2$, pure Weyl-squared, or critical-gravity discussions, where the background and the order of limits matter.

\section{Branch-complete parameter classification}
\label{sec:classification}

\subsection{Regular branches}

\Cref{tab:branches} summarises the five regular structural branches.  ``Healthy'' refers strictly to the quadratic Minkowski spectrum coupled to a conserved metric source.  It does not imply nonlinear consistency or radiative stability.

\begin{table}[!t]
\centering
\caption{Regular branches of the $\Diff$-invariant physical kernel.  The ultraviolet entries describe the source-coupled propagator at large $|q|$.}
\label{tab:branches}
\small
\begin{tabularx}{\textwidth}{@{}p{2.1cm}p{2.4cm}p{3.2cm}c p{2.7cm}X@{}}
\toprule
Branch & Conditions & Particle content & $N_{\DOF}$ & UV behaviour & Health statement for $A>0$ \\
\midrule
Generic & $C\neq0$, $D_s\neq0$, $B\neq2C$ & massless spin-two, massive spin-two, massive scalar & 8 & $k^{-4}$ in both channels & finite massive spin-two pole has opposite residue \\
Mass-degenerate & $B=2C$, $A C\neq0$ & same as generic, equal nonzero masses & 8 & $k^{-4}$ in both channels & two simple eigenspaces, spin-two ghost remains \\
Scalar reduced & $D_s=0$, $C\neq0$ & massless spin-two, massive spin-two & 7 & tensor $k^{-4}$, scalar $k^{-2}$ & massive spin-two ghost remains \\
Spin-two reduced & $C=0$, $B\neq0$ & massless spin-two, massive scalar & 3 & tensor $k^{-2}$, scalar $k^{-4}$ & fully healthy when $B>0$ \\
GR/STEGR & $B=C=0$ & massless graviton & 2 & $k^{-2}$ & fully healthy \\
\bottomrule
\end{tabularx}
\end{table}

The sign-cell analysis refines each structural branch by the signs of $A$, $A/C$, and $A/D_s$.  Positive $A$ fixes the massless graviton orientation.  Positive mass squares remove tachyons but do not repair the massive spin-two residue. Because the massive spin-two residue is opposite to the massless $1/A$ residue independently of $C$, no sign choice can make both spin-two poles healthy; the condition $A/C>0$ independently controls the absence of a spin-two tachyon. This is the local polynomial spin-two obstruction in its branch-resolved form.

\subsection{The two healthy regular loci}

The full regular classification leaves precisely
\begin{align}
\mathcal H_{\mathrm{GR/STEGR}}
&=\{A>0,\ B=0,\ C=0\}
\label{eq:healthyGR}\\
\mathcal H_{\mathrm{scalar}}
&=\{A>0,\ B>0,\ C=0\}
\label{eq:healthyscalar}
\end{align}
The second locus contains the first as its $B\to0$ endpoint only in the sense of a regular infinite-mass limit for the scalar.  Its spectrum consists of the two-helicity graviton and one massive scalar with
\begin{align}
m_s^2=\frac{2A}{3B}>0
\end{align}
This is the unique nontrivial fully healthy regular higher-derivative locus found by the metric free-theory analysis.

The result is both restrictive and useful.  It does not say that every action whose quadratic projection satisfies $C=0$ is a complete quantum theory.  Cubic and quartic vertices can regenerate the excluded spin-two operator through loops unless a symmetry or structural identity protects the locus.  The classification instead identifies the exact classical target that a nonlinear completion must preserve.

\subsection{Ultraviolet behaviour and the classical trade-off}

At large momentum,
\begin{align}
\D_2(q)&\sim
\begin{cases}
q^{-2}& C\neq0\\
q^{-1}& C=0
\end{cases}
&
\D_0(q)&\sim
\begin{cases}
q^{-2}& D_s\neq0\\
q^{-1}& D_s=0
\end{cases}
\end{align}
The generic and equal-mass branches achieve $k^{-4}$ falloff in both physical channels, but they contain the massive spin-two ghost.  The healthy scalar branch improves only the scalar channel; its tensor exchange remains Einstein-like.  GR/STEGR remains $k^{-2}$ in the complete conserved exchange.

This sharpens the usual higher-derivative tension.  In the present finite local metric theory, full two-channel ultraviolet improvement and a healthy regular spin-two spectrum do not coexist.  The statement is deliberately limited to tree-level propagator falloff.  It is not a loop calculation and does not prove or disprove renormalisability of any nonlinear parameter subspace.  It does, however, identify which kinetic terms are available to suppress propagators and which physical poles accompany them.

\section{Conclusion}
\label{sec:conclusion}

We have analysed the classical free theory generated by the complete local parity-even symmetric teleparallel pure-gravity operator basis through four derivatives.  The starting point contains 116 independent nonlinear operators distributed as $5+13+29+69$ among $Q^2$, $(\nabla Q)^2$, $Q^2\nabla Q$ and $Q^4$.  Around Minkowski spacetime in the coincident gauge, perturbative order counting projects this hierarchy onto the $Q^2$ and $(\nabla Q)^2$ sectors.  The explicit operators in Appendix \ref{app:explicit forms of operators} reduce, modulo total derivatives, to four two-derivative and five four-derivative bilinears.  In the convention used here the four-derivative representatives $\mathcal B_1^{(4)}$ and $\mathcal B_4^{(4)}$ carry the minus signs displayed in \Cref{sec:quadratic}; this makes the operator dictionary, the Hessian and the momentum kernel exactly consistent.

The position-space Hessian and independently varied Euler tensor lead to the same momentum operator.  Linearised Diff invariance imposes six independent Ward relations and leaves the three-parameter locus $(A,B,C)$.  The broader quadratic symmetry analysis shows that this is one member of a richer atlas: TDiff has rank two and nullity seven, Weyl has rank four and nullity five, and WTDiff is their exact rank-six intersection with nullity three.  The generic five-generator search further contains accidental double-gradient, mixed-vector, mixed-scalar and derivative-scalar families together with restricted harmonic zero modes.  This atlas is useful because it identifies the field-space degeneracies that must be removed before inversion without confusing coupling-space nullity with physical degree-of-freedom counting.

The Barnes--Rivers representation is a change of basis of the same certified kernel.  On the Diff locus the spin-one and longitudinal scalar coefficients vanish and the physical operator is
\begin{align}
\K_{\Diff}(q)
=q(A+2Cq)\Ptwo
-q\bigl[2A+(3B-2C)q\bigr]\Pzs
\end{align}
A covariant two-parameter gauge fixing renders the full tensor operator invertible, while conserved-source saturation removes every gauge-dependent longitudinal term.  The resulting exchange amplitude is therefore
\begin{align}
\A(q)=
\frac{3I_1-T^2}{3q(A+2Cq)}
-\frac{T^2}{3q\,[2A+(3B-2C)q]}
\end{align}
which is the master expression behind the pole, residue and branch classification.

For $A\neq0$, $C\neq0$ and $3B-2C\neq0$, the regular spectrum contains the two helicities of the massless graviton, five polarisations of a massive spin-two state and one massive scalar, giving eight degrees of freedom.  Their masses are
\begin{align}
m_2^2=\frac{A}{2C}
\qquad\qquad
m_s^2=\frac{2A}{3B-2C}
\end{align}
The spin-two factor admits the exact decomposition
\begin{align}
\frac{1}{q(A+2Cq)}
=\frac1A\left(\frac1q-\frac1{q+m_2^2}\right)
\end{align}
so once the massless graviton orientation is fixed by $A>0$, every finite massive spin-two pole has the opposite residue.  No scalar tuning repairs this sign opposition.  Equal nonzero masses at $B=2C$ do not create a double pole because the spin-two and scalar factors multiply orthogonal eigenspaces.  Critical $A=0$ surfaces and identically vanishing kinetic blocks instead change the analytic order or rank of the operator and must be treated separately from ordinary particle branches.

The complete regular classification leaves precisely two fully healthy Minkowski loci for a conserved metric source
\begin{align}
\mathcal H_{\mathrm{GR/STEGR}}&=\{A>0\quad B=0\quad C=0\}\\
\mathcal H_{\mathrm{scalar}}&=\{A>0\quad B>0\quad C=0\}
\end{align}
The first carries only the two graviton helicities.  The second carries the massless graviton plus one non-tachyonic massive scalar with $m_s^2=2A/(3B)$ and therefore three degrees of freedom.  The scalar-reduced branch retains the massive spin-two ghost, while the generic and equal-mass branches achieve $k^{-4}$ suppression in both source-coupled channels only at the price of that ghost.  The healthy scalar branch improves only the scalar channel, leaving the tensor exchange at $k^{-2}$.  Thus in the finite local metric theory full four-derivative tensor suppression and a healthy regular spin-two spectrum do not coexist.

These statements reproduce the standard curvature-squared benchmarks after convention translation.  GR/STEGR lie at $B=C=0$ and $R+R^2$ lies on the $C=0$ scalar branch. Einstein--Weyl gravity removes the scalar but retains the massive spin-two ghost and generic quadratic gravity gives the familiar eight-degree-of-freedom spectrum with improved ultraviolet falloff and the Stelle ghost \cite{Stelle1977,Stelle1978,Starobinsky1980,Salvio2018,LuPerkinsPopeStelle2015}.  Pure $R^2$, pure Weyl-squared and critical-gravity limits belong to singular or background-sensitive sectors and cannot be inferred by naively taking regular pole masses to zero \cite{LuPope2011}.

Within symmetric teleparallel gravity, the metric kernel maps exactly to the affine-polynomial specialisation of the Conroy--Koivisto form factors
\begin{align}
\mathfrak a(\Box)=A-2C\Box
\qquad\qquad
\mathfrak c(\Box)=A-B\Box
\end{align}
so that the present result agrees with their general analytic propagator while adding the explicit finite operator origin and branch-complete local classification \cite{ConroyKoivisto2018}.  The broader symmetric-teleparallel literature developed by Adak, Dereli, Pala and collaborators supplies complementary formulations, exact solutions, gauge-theoretic treatments and geometric analyses of nonmetricity gravity \cite{AdakSert2005,Adak2006,AdakDereli2008,Adak2018,PalaAdak2022,PalaSertAdak2023,AdakDereliKoivistoPala2023,AdakOzdemirPala2023}.  Our result is narrower: it isolates the free metric spectrum of the complete local four-derivative operator hierarchy around Minkowski spacetime.

Analytic $f(Q)$ gravity provides an instructive boundary case.  Since the nonmetricity scalar itself begins at $O(h^2)$ around a $Q=0$ Minkowski vacuum, only $f_Q(0)$ enters the quadratic action when $f_Q(0)\neq0$.  The corresponding free metric kernel is therefore GR/STEGR-like, while additional modes reported in cosmological or Hamiltonian analyses probe nonlinear, non-Minkowski or strongly coupled sectors that are not visible in this regular quadratic limit \cite{JimenezHeisenbergKoivistoPekar2020,Hu2022,Dambrosio2023,Gomes2024,HuYamakoshi2023}.

The same transferability boundary applies to enlarged metric-affine and Riemann--Cartan models.  Healthy massive spin-two, spin-three, scalar or vector states found in those theories can live partly or wholly in independent connection, distortion or torsion sectors \cite{BaldazziMelichevPercacci2022,PercacciSezgin2020,MikuraPercacci2025,Percacci2026,AokiMukohyama2019,Neville1978,SezginVanNieuwenhuizen1980,Sezgin1981,LinHobsonLasenby2019,Marzo2022}.  They are therefore not counterexamples to the metric-only residue theorem derived here.  A direct comparison requires the same independent fields, geometric constraints, gauge quotient and source-coupled quadratic operator.

The present analysis consequently fixes the classical target for the nonlinear stage of this work.  Distinct nonlinear actions can project onto the same nine-dimensional free theory, and the healthy condition $C=0$ need not be radiatively stable without additional structure.  The $Q^2\nabla Q$ and $Q^4$ sectors determine cubic and quartic vertices, while the full $Q^2$ and $(\nabla Q)^2$ sectors provide the available quadratic counterterms.  The next questions are therefore whether the nonlinear coupling subspace projecting onto the healthy scalar branch is preserved by the interaction algebra, whether loop corrections regenerate the excluded spin-two operator and how the free-theory trade-off changes once the full interacting and quantum problem is addressed.  Until those calculations are performed, the conclusions above should be read as a branch-complete classical free-spectrum result rather than a claim of perturbative renormalisability or nonlinear viability.

\section*{Acknowledgements}
During the preparation of this manuscript, the authors used artificial intelligence (AI)-based tools as assistive resources for language editing, brainstorming, cross-checking results and calculations, and coding. All AI-generated or AI-assisted content incorporated into the work was critically reviewed, verified, and, where appropriate, revised by the authors. The authors take full responsibility for the accuracy, integrity, interpretation, and originality of the manuscript and its results.

\newpage
\appendix
\section*{Appendices}
\renewcommand{\thesubsection}{\Alph{subsection}}
\setcounter{subsection}{0}

\subsection{Explicit forms of $Q^2$ and $(\nabla Q)^2$ sectors of the Lagrangian}\label{app:explicit forms of operators} 
Here we display explicitly the two free sectors of the theory appearing in Eq.\eqref{eq:Lagrangian original free theory 18 terms}. The sector quadratic in nonmetricity is
\begin{align}
    \Lag_{Q^2}= a_1 Q_{\alpha\mu\nu}Q^{\alpha\mu\nu}
+a_2 Q_{\alpha\mu\nu}Q^{\mu\alpha\nu}
+a_3 Q_\alpha Q^\alpha
+a_4 Q_\alpha\widetilde Q^\alpha
+a_5 \widetilde Q_\alpha\widetilde Q^\alpha
\end{align}
while the sector quadratic in derivatives of nonmetricity is
\begin{align}
    \Lag_{(\nabla Q)^2} &= b_1\,(\partial_\mu Q^\mu)^2 + b_2\,\partial_\mu Q^\mu \partial_\nu\widetilde{Q}^\nu + b_3\,\partial_\rho Q^\rho{}_{\mu\nu}\partial^\mu Q^\nu + b_4\, \partial_\rho Q^\rho{}_{\mu\nu} \partial^\mu\widetilde{Q}^\nu + b_5\, \partial_\rho Q^\rho{}_{\mu\nu} \partial_\sigma Q^{\sigma\mu\nu} \nonumber\\
&\quad + b_6\, \partial_\mu\widetilde{Q}^\mu \partial_\nu\widetilde{Q}^\nu + b_7\, \partial_\rho Q_{\mu\nu}{}^\rho \partial^\mu Q^\nu + b_8\, \partial_\rho Q_{\mu\nu}{}^\rho \partial^\mu \widetilde{Q}^\nu + b_9\, \partial_\rho Q_{\mu\nu}{}^\rho \partial^\nu \widetilde{Q}^\mu + b_{10}\, \partial_\mu Q_\nu \partial^\mu Q^\nu \nonumber\\
&\quad + b_{11}\, \partial_\rho Q_{\sigma\mu\nu} \partial^\rho Q^{\sigma\mu\nu} + b_{12}\, \partial_\rho Q_{\sigma\mu\nu} \partial^\rho Q^{\mu\sigma\nu} + b_{13}\, \partial_\rho Q_{\sigma\mu\nu} \partial^\mu Q^{\nu\rho\sigma}
\end{align}

\subsection{Conditions on the bilinear coefficients for the symmetries examined in Sec. \ref{sec:reduced symmetries}}\label{app:symmetry conditions}
Diff, TDiff, Weyl and WTDiff symmetries and their corresponding coupling conditions have already been discussed in the main text. In this appendix we collect the coupling conditions for the principal additional branches; exceptional charts and restricted images are summarised in Table \ref{tab:symmetry-atlas}.
\begin{enumerate}
    \item $V_2$ branch : Trace image of Weyl
    \begin{subequations}
        \begin{align}
        \alpha_1+8\alpha_2+2\alpha_3&=0 &\qquad \beta_1+8\beta_2+2\beta_3 &=0 \\
        2\alpha_1+\alpha_4&=0 &\qquad 2\beta_1+\beta_4+\beta_5 &=0
    \end{align}
    \end{subequations}
    \item $S_1$ branch : Double gradient
    \begin{subequations}
            \begin{align}
        \alpha_1+2\alpha_2 &=0 &\qquad \beta_1+2\beta_2 &=0 \\
        \alpha_1+2\alpha_3+2\alpha_4 &=0 &\qquad \beta_1+2\beta_3+2\beta_4+2\beta_5 &=0
    \end{align}
    \end{subequations}
    \item $S_2$ branch : Metric-box image\\
    At generic nonzero momentum this branch has the same coupling ideal as Weyl, since $S_2=qS_0$.
    \item $V_1+rV_2$ branch : Mixed vector (general)
    \begin{subequations}
        \begin{align}
            2\alpha_3+\alpha_4 &= 0 &\qquad 2\beta_3+\beta_4 &=0 \\
            (1+r)\alpha_1+(2+8r)\alpha_2+2r\alpha_3 &= 0 &\qquad (1+r)\beta_1+(2+8r)\beta_2+2r\beta_3 &= 0 \\
            (1+4r)\alpha_1+(1+2r)\alpha_4 &=0 &\qquad (1+4r)\beta_1+(1+2r)\beta_4+2(1+r)\beta_5 &=0
        \end{align}
    \end{subequations}
    \item $S_0+pS_1+tS_2$ branch : Mixed scalar ($p\neq 0$)
    \begin{subequations}
        \begin{align}
            2\alpha_1+\alpha_4 &=0 \\
            \alpha_1+8\alpha_2+2\alpha_3 &=0 \\
            (p+4t)\alpha_1+2p\alpha_3+2(p+t)\alpha_4+4\beta_1+2\beta_4+2\beta_5 &=0 \\
            (p+4t)\beta_1+2p\beta_3+2(p+t)\beta_4+2(p+t)\beta_5 &=0 \\
            (p+t)\alpha_1+2(p+4t)\alpha_2+2t\alpha_3+\beta_1+8\beta_2+2\beta_3 &=0 \\
            (p+t)\beta_1+2(p+4t)\beta_2+2t\beta_3 &=0
        \end{align}
    \end{subequations}
    \item $S_1+tS_2$ branch : Derivative scalar
    \begin{subequations}
        \begin{align}
            (1+t)\alpha_1+(2+8t)\alpha_2+2t\alpha_3 &=0 &\qquad (1+t)\beta_1+(2+8t)\beta_2+2t\beta_3 &=0 \\
            (1+4t)\alpha_1+2\alpha_3+2(1+t)\alpha_4 &=0 &\qquad (1+4t)\beta_1+2\beta_3+2(1+t)(\beta_4+\beta_5) &=0
        \end{align}
    \end{subequations}
    \item $S_0$ branch : Harmonic Weyl ($q\phi=0$)
    \begin{align}
        2\alpha_1+\alpha_4 =0
    \end{align}

\end{enumerate}

\printbibliography

\end{document}